\documentclass{emulateapj}
\usepackage{graphicx}
\usepackage{bm}
\usepackage{amsmath}
\usepackage{natbib}

\newcommand{\eqb}{\begin{eqnarray}}
\newcommand{\eqe}{\end{eqnarray}}

 \slugcomment{} \shorttitle{The long-term evolution of
the GW170817 remnant}
\begin{document}
\title{The long-term evolution of the GW170817 remnant}
\author{Menquan Liu\footnote{corresponding author
} Jie Zhang and Cong Wang} \affil{Department of Astronomy, Qilu
Normal University, Jinan 250300, China} \email{menquan@qlnu.edu.cn}

\begin{abstract}
  GW170817 represents the first observed binary neutron star merger event by humanity.
   The observation of GW170817 has identified the correlation between Kilonova,
   gravitational wave and short GRB. The shocks from GW170817 have the capacity to inject significant
   thermal and kinetic energies into the interstellar medium and evolve for
   over a million years. In this letter, we adopt the special relativity fluid dynamics equations
   to simulate the evolution of the GW170817 remnant over a span of one million years.
  Our simulations yield the evolution profiles of the velocity, density, mass, radius, luminosity,
   and energies of the remnant.  We estimate that the GW170817 remnant will reach the average maximum
luminosity  $ 2.56\times 10^{39}$ erg s$^{-1}$at approximately $
3.96\times 10^4$ yr. At the end of the cooling stage, the
contaminated radius and mass are $48.35$ pc and $2.25\times 10^4
   M_{\odot}$, respectively.
\end{abstract}


\section{Introduction}
On August 17, 2017, the LIGO and Virgo collaborations detected
gravitational waves from the binary neutron star merger source
GW170817 for the first time, and 10.86 hours later, \citet{dro17}
identified signals in the optical bands, including ultraviolet,
optical, and infrared (its transient electromagnetic counterpart is
named AT2017gfo). The analysis indicates that the optical curve is
largely consistent with the decay of the r-process nucleosynthesis
products, and the mass of the eject\textbf{a} during the merger
process is estimated to be $\sim0.05$ $M_{\odot}$. \citet{abb17a}
inferred from the corresponding light curves that the distributions
of dynamical ejecta mass vary between $10^{-3}-10^{-2} M_{\odot}$
for various equations of state. A more recent estimation of ejecta
mass range \textbf{is} between $10^{-3}-10^{-1}
M_{\odot}$\citep{ris23}. The optical and near-infrared observations
are effectively explained using the kilonova model. Furthermore,
emissions in other bands of electromagnetic signals from GW170817,
including X-ray and gamma-ray bands, have also been observed. Thus,
the co-occurrence of gravitational waves and electromagnetic signals
during the merger of binary neutron stars is unequivocally
confirmed. It is hypothesized that the short gamma burst coincides
with the binary neutron star merger event. \citet{kas17} examined
the signals in the gamma band of GW170817/GRB170817a and concluded
that GW170817/GRB170817a is not an ordinary short gamma burst with
an extreme relativistic jet, but a wide-angled, moderately
relativistic, cocoon-like explosion, whose remnant's shape assumes
nearly spherical symmetry over long-term evolution. Recent efforts
to simulate the generation of optical counterparts from the gamma
band to the radio band of GW170817 have been undertaken by two- and
three-dimensional models\citep{eka23}.

Binary neutron star merger events are considered significant sources
of the r-process elements. Current research posits that most of the
r-process elements with $A>130$ in the universe may have originated
from neutron star mergers, rather than supernova explosions as
previously assumed. Compared with neutron star mergers,
core-collapse supernovae exhibit a higher explosion rate, but the
r-process element yield is comparatively low, and more importantly,
they struggle to produce the element distributions for $A>130$ that
align with observations. In addition, neutron star mergers are also
tied to cutting-edge research in astronomical and theoretical
physics, such as dark energy and general relativity\citep{cre17}.
GW170817 may have harbored a magnetar at its center before the
explosion, and a black hole may have formed at its center after the
explosion. However, it was suggested that it was possible that the
merger of strange stars rather than neutron stars
occurred\citep{lai18}.

In summary, GW170817 is a pivotal and typical source, and it is also
one of the hot spots in astronomical and theoretical physics
research. Although recent observations at 3 GHz band indicated that
there is no clear evidence for a late-time re-brightening of the
GW170817 nonthermal afterglow emission, the remnant may be observed
by multiple bands later\citep{bal22}. Present researches on GW170817
mainly report on observed phenomena and how they are interpreted,
and seldom focus on its subsequent evolution of remnant. A study was
conducted on the binary neutron star merger whose evolution time is
more than 100 years\citep{ros14}. The long-term evolution of binary
neutron star merger remnants for millions of years has been
previously explored, but it is not targeted to
GW170817\citep{liu17}. The difference in the long-term evolution of
neutron stars merger and supernova has also been discussed. In this
letter, we aim to substitute the observational parameters of
GW170817 into a non-homogeneous special relativistic hydrodynamic
equations including electron fraction with r-process elements and
the typical interstellar medium properties to simulate the
subsequent long-term evolution of the GW170817. Given the long-term
evolution, ejecta mass is more important than other quantities such
as precise shape, composition, temperature and density of the
ejecta. Therefore, we will examine the effect of different ejecta
mass on the long-term evolution. This is a novel approach. Our study
can predict not only the subsequent evolution of the GW170817, but
also the impact on the interstellar medium around it. The evolution
laws of  luminosity, energies, velocity, etc., of the remnant will
be detailed.

\section{Hydrodynamical Equations and Initial Model }
\label{twofluid}

Due to the substantial kinetic energy of the ejecta coupled with its
minimal mass, the velocity achieved is remarkably high.
Specifically, for an ejecta mass of $0.001M_{\odot}$, the average
speed reaches approximately $\sim0.77c$, where $c$ is the velocity
of light. Under these conditions, the traditional Newtonian dynamics
equations fail to provide an accurate representation, thereby
necessitating the adoption of special relativistic hydrodynamic
equations.

The special relativistic hydrodynamic equations, incorporating
cooling mechanisms for the detailed depiction of shock propagation
in remnants, are articulated as follows \citep{liu17}

 \begin{eqnarray}
\frac{\partial D}{\partial t}+\nabla \cdot (D \textbf{\emph{v}})=0,\label{eqms}\\
\frac{\partial \textbf{\emph{S}}}{\partial t}+\nabla \cdot (\textbf{\emph{S}}\textbf{\emph{v}}+P\emph{I})=0,\label{eqmo} \\
\frac{\partial{\tau}}{\partial {t}}+\nabla
\cdot(\tau\textbf{\emph{v}}+P\textbf{\emph{v}})=-n_en_H\Lambda(T),\\
\textbf{\emph{S}}=Dh\Gamma\textbf{\emph{v}}, \\
\tau=Dh\Gamma-P-D,
 \end{eqnarray}
 where Eqs.(\ref{eqms})-(3) represent the conservation of rest mass, momentum,
 and energy, respectively. The variables $D$, $\textbf{\emph{S}}$ and $\tau$ correspond
 to the lab frame rest mass, momentum, and energy densities (excluding rest mass),
 respectively. $\textbf{\emph{D}}=\rho \Gamma$, where $\rho$ is the mass density,
 $\Gamma=(1-v^2)^{-1/2}$ is the Lorentz factor. $h$ is the specific enthalpy defined
 by $h=1+\epsilon+P/\rho$, with $\epsilon$ being the specific internal energy.
  $P$ is the pressure. In the above equations, we have used the natural units
  in which the speed of light $c$ \textbf{is} taken to be unity.
  For example, $h=1+\epsilon/c^2+P/\rho c^2$  in the normal units.
  In the equations, $t$ is the time, $\textbf{\emph{v}}$ is the velocity and $\emph{I}$
  is the identity matrix. $n_e$ and $n_H$ are the electron and the hydrogen number densities,
   respectively. $\Lambda(T)$ is the cooling function.  For temperatures above $10^4$K,
   the cooling function is given by \citet{wan14}. While for temperatures between $10^2-10^4$ K,
    the cooling function by \citet{sar22} is adopted. As the temperatures \textbf{drop} below $10^2$ K,
    the cooling rate is not effective and \textbf{is} treated as zero. Assuming
    the composition of interstellar medium(ISM) is monatomic and nonrelativistic,

\begin{equation}
 P=\rho\mu^{-1}N_A kT,
\end{equation}
 where $N_A$ represents the Avogadro constant,  $k$ the Boltzmann constant,
 and $\mu$ the mean molecular weight in units of $m_{\rm H}$.
 In the case of ionized gas, $\mu$ is determined by following equation
\begin{equation}\mu^{-1}
=Y_e+Y_N, \end{equation}
\begin{equation}Y_e
=X_H+1/2X_{He}+\sum_{i}\frac{Z_i}{A_i}X_{Ri},
\end{equation}
\begin{equation}Y_N
=X_H+1/4X_{He}+\sum_{i}\frac{1}{A_i}X_{Ri}, \end{equation}
 where $Y_e$ and $Y_N$ are electron fraction and nucleus fraction
 for each nucleon. $X_H$ and $X_{He}$ are the mass abundance for hydrogen and
 helium, respectively. $Z_i$, $A_i$ and $X_{Ri}$  are the nuclear charge number,
 mass number, and mass abundance of metals (including r-process
 elements).

 The methodology employed to solve the above equations is an explicit
 Lagrangian finite-difference scheme as we employed
 before\citep{liu09,liu17}.  The explosive energy of GW170817 is
 $\sim 1.0\times10^{51}$ erg according to \citet{shi17}, in which
 kinetic energy is the vast majority, and thermal energy is
 negligible. The initial $Ye$ of ejecta is 0.25, but after decay of
 r-process elements, it changes to 0.47. The composition of ISM is the
 recommended elemental abundances in the solar photosphere derived
 from spectroscopy. The density of ISM is $2.22\times10^{-24}$ g
 cm$^3$(corresponding to a number density of hydrogen $n_H=1$ cm$^3$\textbf{)}. The
 ISM temperature is set to be 10$^3$ K to make sure it doesn't cool
 down until the shock arrives. Since the maximal uncertainty comes
 from ejecta mass and initial velocity, we simulated three different  evolution models with ejecta masses as follows:
   $10^{-3}M_{\odot}$(noted as Model 1) \citep{rue22,ris23}, $0.05 M_{\odot}$ (noted as Model 2) \citep{dro17} and $10^{-1}M_{\odot}$ (noted as Model 3) \citep{fry24}. $0.05 M_{\odot}$ is the most likely ejecta mass of GW170817, while $10^{-3}M_{\odot}$  and $10^{-1}M_{\odot}$  represent the extremes of
 ejecta masses. Their parameters are shown in the Table 1.

\section{Results }
The evolution of binary neutron star merger remnants can be divided
into three stages, the first of which is the adiabatic expansion
stage before the shock wave cools. The second stage is the cooling
phase. In this stage, a large amount of energy is emitted, causing
the luminosity of the remnant to significantly increase. On the
other hand, a relatively dense shell is formed rapidly at the same
time. When the total energy released by cooling reaches 90\% of the
initial energy, we consider the cooling to be over and the evolution
enters the third stage, namely, the shell dispersion stage, in which
the thickness of the shell increases significantly, and the density
becomes smaller and smaller, gradually approaching the ISM density.

The evolution of the three models is similar, so in the following we
will focus on the most possible initial model: Model 2. Figure 1
shows the velocity profiles at some characteristic times, ranging
from $10^2$ yr to $10^6$ yr after the merger,
in which $3.55\times10^4$ year is the time when the remnant reaches
the maximum luminosity. From the velocity profile of the shock, it
is evident that the wave front of the shock is very steep in the
first stage, until the relatively dense shell begins to form. The
first shock wave is the most important part of the shock wave
propagation process because it contains the main mass and energy of
the remnant. We present the variation of the velocity of the first
shock wave in the bottom panel of Figure 1. It can be seen that the
speed of the first shock wave decreases very quickly, and the speed
of the shock wave generally decreases exponentially. However, it
does not decrease monotonically; we can see that the speed of the
shock wave first decreases and then increases at $\sim 3.6\times
10^4$ yr, forming a V-shape, attributed to the deceleration of the
first shock when the cooling begins, until the secondary shock
(sub-shock) catches up with the first shock wave. Shock wave piles
up ISM into a shell, and the area behind the shell is called as
\lq\lq bubble\rq\rq. The change of velocity inside the bubble is
complex. In the initial stage, the velocity inside the bubble
gradually decreases with time, and then an inward shock wave is
formed. The inward shock wave collides with the remnant center and
bounces back. The bounce shock moves forward fast and can catch up
with the first shock. This shock oscillation occurs several times in
the bubble, but the amplitude of the oscillation is getting smaller
and smaller with time. After $10^6$ yr, the speed is  less than 20
km s$^{-1}$.

\begin{table*}
\centering
 \caption{Evolutionary parameters at different ejecta masses, where $M_{\rm{ej}}$ is the ejecta mass, $\bar{v}_{\rm{ini}}(c)$
 is the initial average velocity. The subscript $m$ indicates the time when the luminosity is maximal. The subscript $E$
 indicates the end time for cooling.}
  \renewcommand{\arraystretch}{1.5}
  \setlength{\tabcolsep}{0.8mm}{ 
\begin{tabular}{l l c c c c c c c c}\hline
$$&$M_{\rm{ej}}(M_{\odot})$ & $\bar{v}_{\rm{ini}}(c)$& $t_m$(yr) & $R_m
$(pc) & $L_m(\rm{erg \,s}^{-1}) $ &
$t_E(\rm{yr}) $ & $R_E $(pc) & M$_E(M_{\odot}$)\\
\hline${\rm Model\, 1}$&$1.0\times10^{-3}$& 0.77 & $3.31\times10^4$& 18.84& $1.04\times 10^{39}$&$4.79\times10^5$&41.41 &$9.71\times10^3$\\
${\rm Model\, 2}$&$5.0\times10^{-2}$&0.32 & 3.55$\times 10^4$&20.69&1.31$\times 10^{39}$&$4.90\times10^{5}$& 44.74&$1.22\times10^4$ \\
${\rm Model\, 3}$& $1.0\times10^{-1}$&0.22& 5.01$\times10^4$& 31.99&$4.08\times10^{39}$& $6.92\times10^5$&69.30& $4.55\times10^4$ \\
\hline
\end{tabular}}
\label{tabpara}
\end{table*}

Figure 2 shows the density profile of the remnant over a million
years. The characteristic times are the same as those in Figure 1.
After the adiabatic expansion stage, the maximum luminosity appears
when the shell is first formed, because the temperature of the shell
at this time is about $10^5$ K, at which the value of the cooling
function is maximal. The effective cooling converts the heat energy
of the remnant into light energy and \textbf{is} released outward.
It is evident that during the cooling stage, the shock wave piles up
the ISM into a very thin shell, as illustrated by the density
profile at $10^5$ yr, which is more than 3 orders of magnitude
denser than the average density of the ISM. On the contrary, the
density in the bubble is very low, similar to a vacuum. There is
another shell inside the bubble resulting from the boundary of the
ejecta and the initial ISM. At the third stage, the shell slowly
disappears and spreads, as shown in the density profile at $10^6$yr.
This is the final stage of the remnant evolution.

The mass and radius of the contaminated ISM vs time are shown in
Figure 3. Understandably, both quantities exhibit a monotonic
increase over time. It can be observed that the variation of the
radius and mass of the remnant is relatively smooth. The radius and
mass of the remnant at the time of maximum luminosity are 20.69 pc
and $1.20\times 10^3 M_{\odot}$, respectively. However, the radius
and mass of the remnant increase to 44.74 pc and $1.22\times 10^4
M_{\odot}$ after the cooling stage. The contaminated ISM mass and
radius are significantly influenced by the ejected energy and mass
(see Table 1).

Energy and luminosity are complementary concepts, as depicted in
Figure 4 and Figure 5. Figure 4 presents the changes in kinetic
energy, thermal energy, and total energy of the remnant at different
times. The black, red, and blue lines represent the kinetic,
thermal, and total energy of the remnant, respectively. Although the
total energy remains essentially constant during the adiabatic
expansion stage, resulting in a nearly horizontal trend, the
composition of energy undergoes significant transformations,
shifting from predominantly kinetic to thermal energy dominance at
this stage. During the cooling stage, kinetic energy, thermal
energy, and total energy all decline sharply. In the diffusion
phase, the overall energy is dissipated gradually. The adiabatic
expansion period lasts for $1.2\times10^4$ years, with the maximum
luminosity occurring at $3.55\times10^4$ years, reaching
$1.31\times10^{39}$ erg s$^{-1}$, and the luminosity remains larger
than $10^{38}$ erg s$^{-1}$ from $1.86\times 10^4$ yr to $8.13\times
10^4$ yr, spanning about $6.3\times 10^4$ years. Post-peak
luminosity, the trend exhibits significant oscillations alongside a
general exponential decrease. Compared to supernova remnants, the
energy variation trend of the merger remnant's shock wave closely
mirrors that of supernova remnants, with the notable difference
being that the onset of cooling for the merger remnant (marked by
the peak luminosity) occurs significantly earlier than in supernova
remnants. In Figure 5, it is evident that the greater the mass of
the ejecta, the higher the luminosity, and the longer the duration
before reaching the maximum luminosity.

\section{conclusion and discussion}

From the above analysis, we estimate that the variation range of
GW170817 remnant. For example, the time to reach the maximum
luminosity is $3.31\sim 5.01\times 10^4$ yr. The maximum luminosity
is $1.04\sim 4.08\times 10^{39}$ erg s$^{-1}$. Taking the average,
we estimate that the GW170817 remnant will reach the maximum
luminosity at approximately $\sim 3.96\times 10^4$ yr, with the
maximum luminosity being around $2.56\times 10^{39}$ erg s$^{-1}$.
The remnant radius and mass when the cooling ends are estimated to
be 48.35 pc and $2.25\times 10^4 M_{\odot}$, respectively.

The metallicity of the surrounding ISM of a binary neutron star may
be greater than that of the Sun, thus all the aforementioned
quantities will vary with the metallicity. According to the
empirical formula\citep{tho98}, the remnant radius $R_E$ is
proportional to $(Z/Z_{\odot})^{-0.1}$, and the remnant mass $M_E$
is proportional to $(Z/Z_{\odot})^{-0.28}$. If the actual
metallicity is twice that of the Sun, the remnant radius and mass
upon the completion of cooling would be 51.82 pc and $1.85\times
10^4 M_{\odot}$, respectively. This principle applies similarly to
other quantities.

\begin{acknowledgements}
 MQL like to thank professor Y. Z. Qian at the University of Minnesota for his help.
This work is supported by the National Natural Science Foundation of
China (grant 12373109 and U2031121), National Key Research and
Development Program of China (grant SQ2023YFB4500096) and the
Natural Science Foundation of Shandong Province (grant ZR2020MA063).
\end{acknowledgements}

\bibliographystyle{hapj}
\bibliography{NSM}

\begin{figure*}
\centering
\begin{tabular}{c}
\includegraphics[scale=0.8]{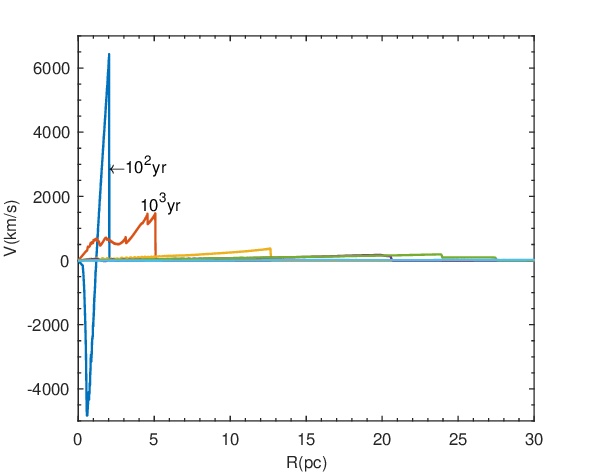}\\
\includegraphics[scale=0.8]{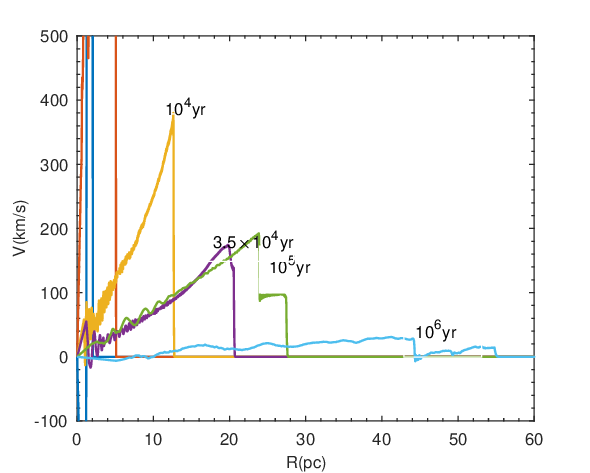}\\
\includegraphics[scale=0.4]{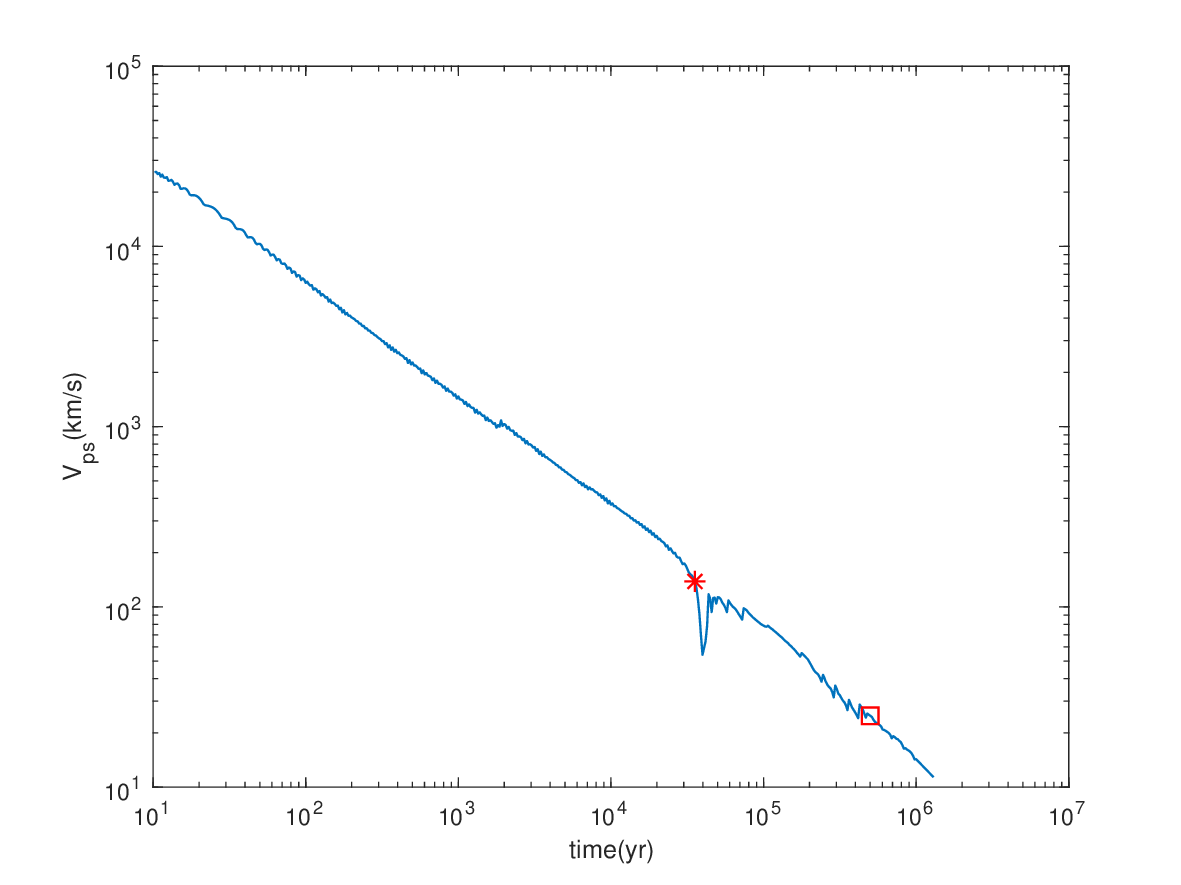}
\end{tabular}
\caption{Subsequent velocity evolution of the GW170817 remanant.The
top panel is the velocity profiles at $10^2$ and $10^3$ yr, the
middle panel is the velocity profiles at $10^4, 3.55\times 10^4,
10^5$ and $10^6$ yr, the bottom panel is the velocity of the first
shock vs time, in which star and square denote the time of the
maximum luminosity and the end of cooling.} \label{fig1}
\end{figure*}

\begin{figure*}
\centering
\includegraphics[scale=0.4]{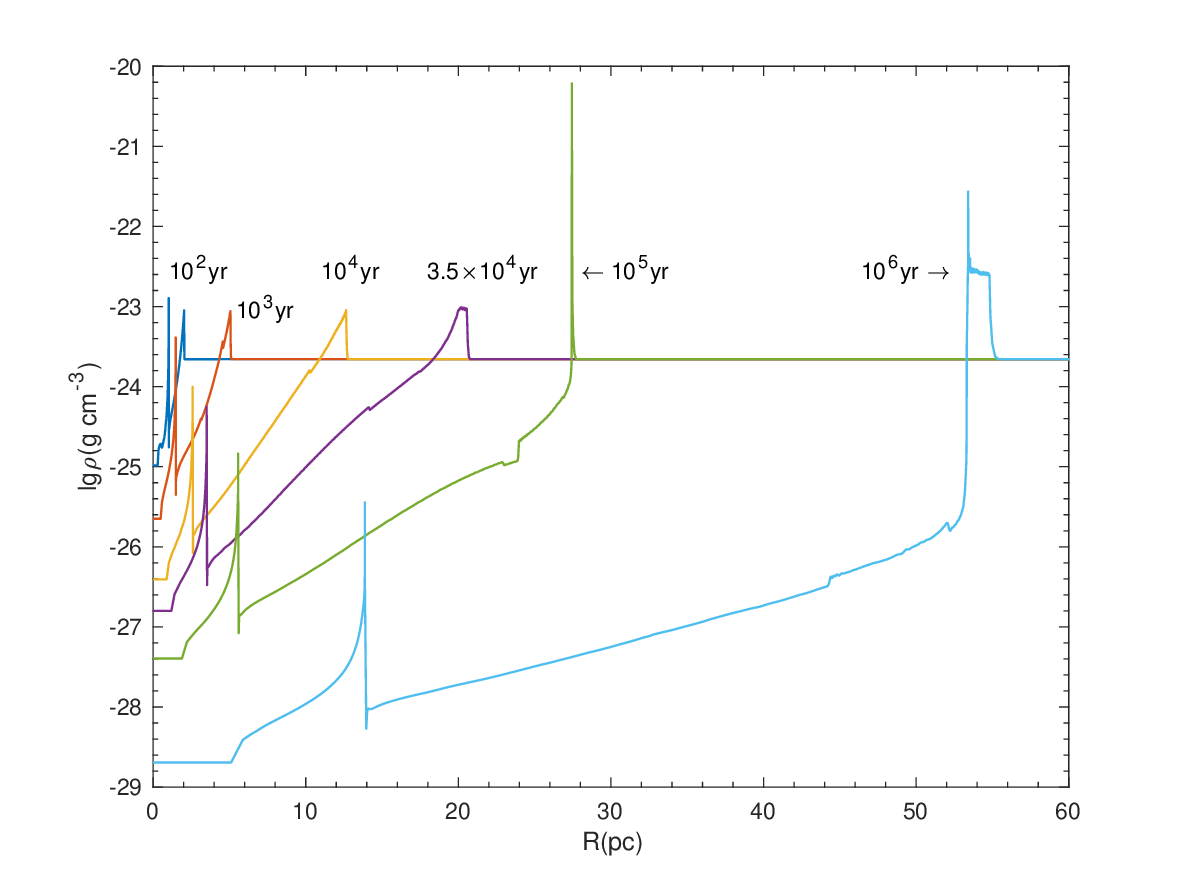}
\caption{Density profiles of the GW170817 remnant at different time.
The characteristic time are the same as that in Fig.1. }
\label{fig2}
\end{figure*}

\begin{figure}
\centering
\begin{tabular}{c}
\includegraphics[scale=0.4]{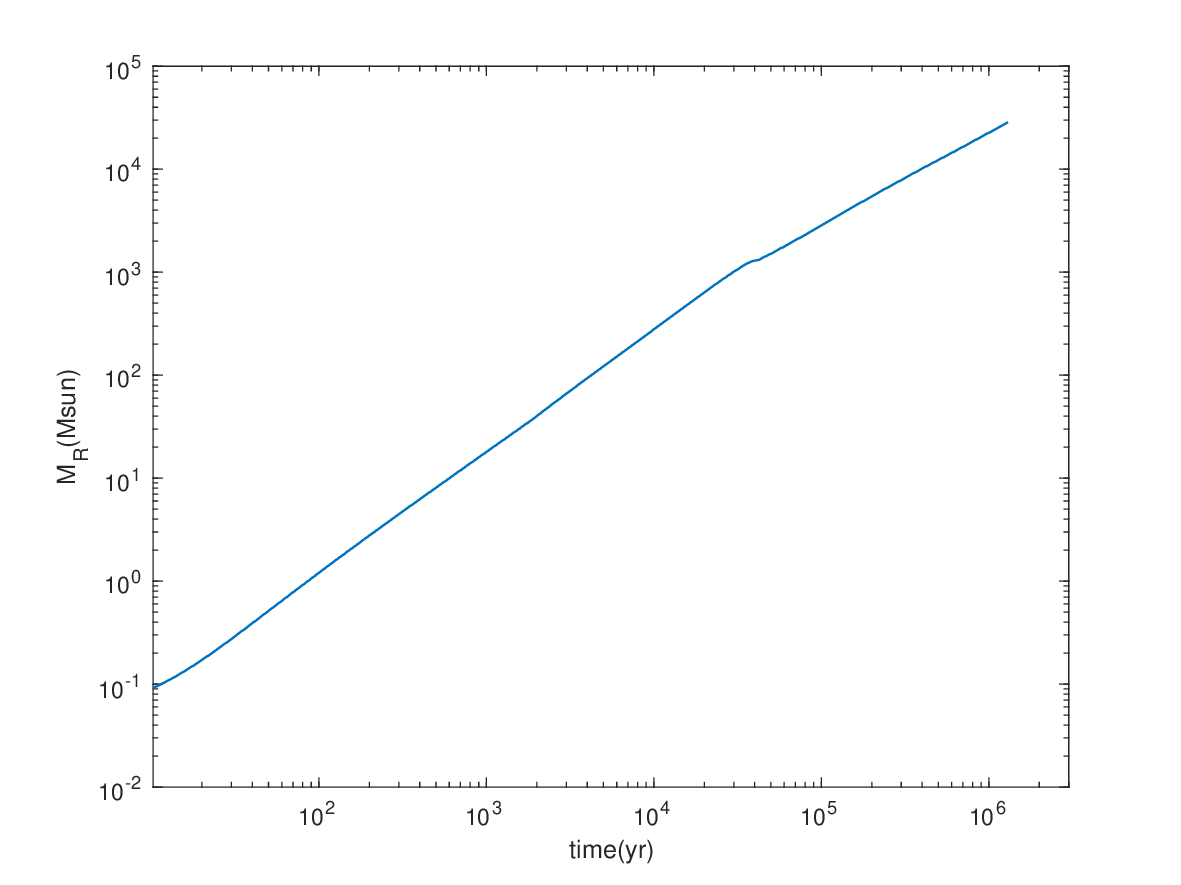}\\
\includegraphics[scale=0.4]{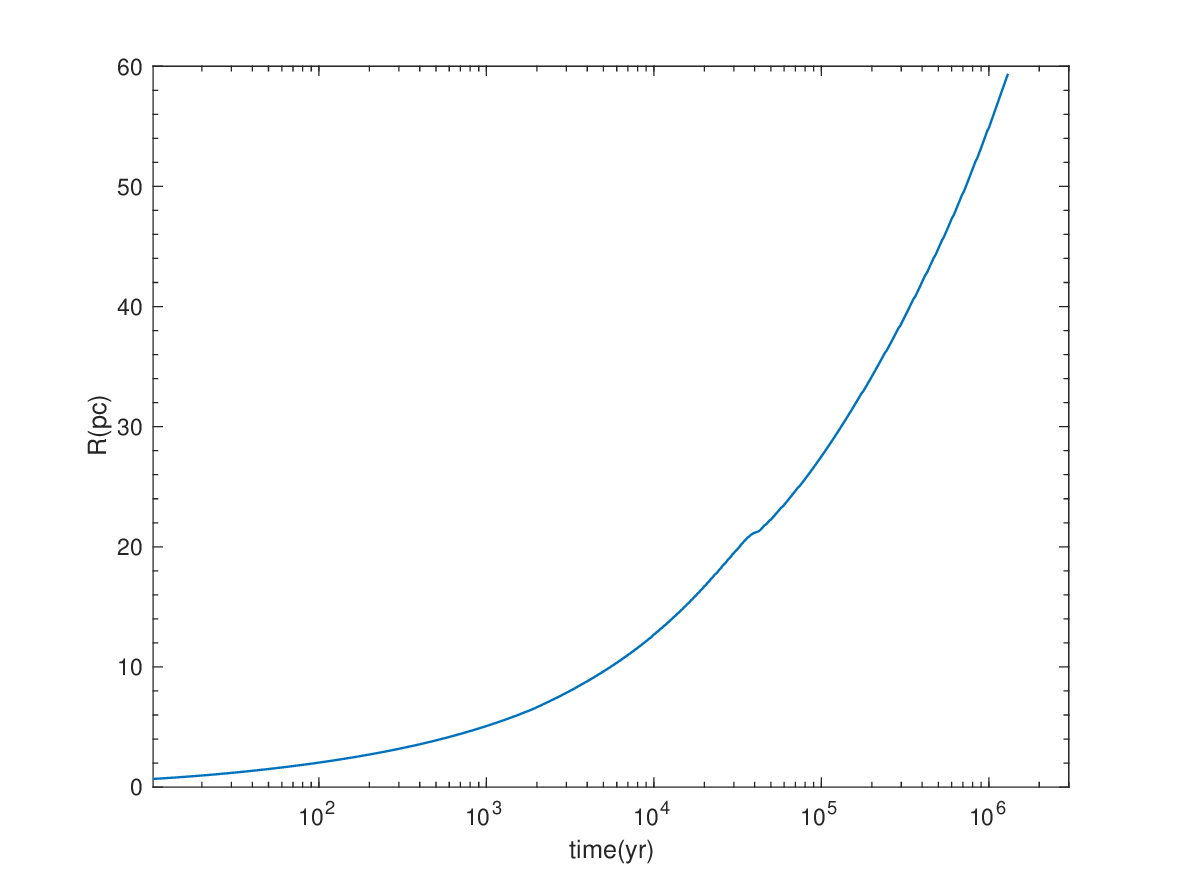}
\end{tabular}
\caption{The mass (top) and radius (bottom) of the contaminated ISM
as a function of time.The small circles and small squares represent
the time when the luminosity is maximal and when the cooling ends,
respectively.} \label{fig3}
\end{figure}

\begin{figure}
\centering
\includegraphics[scale=0.3]{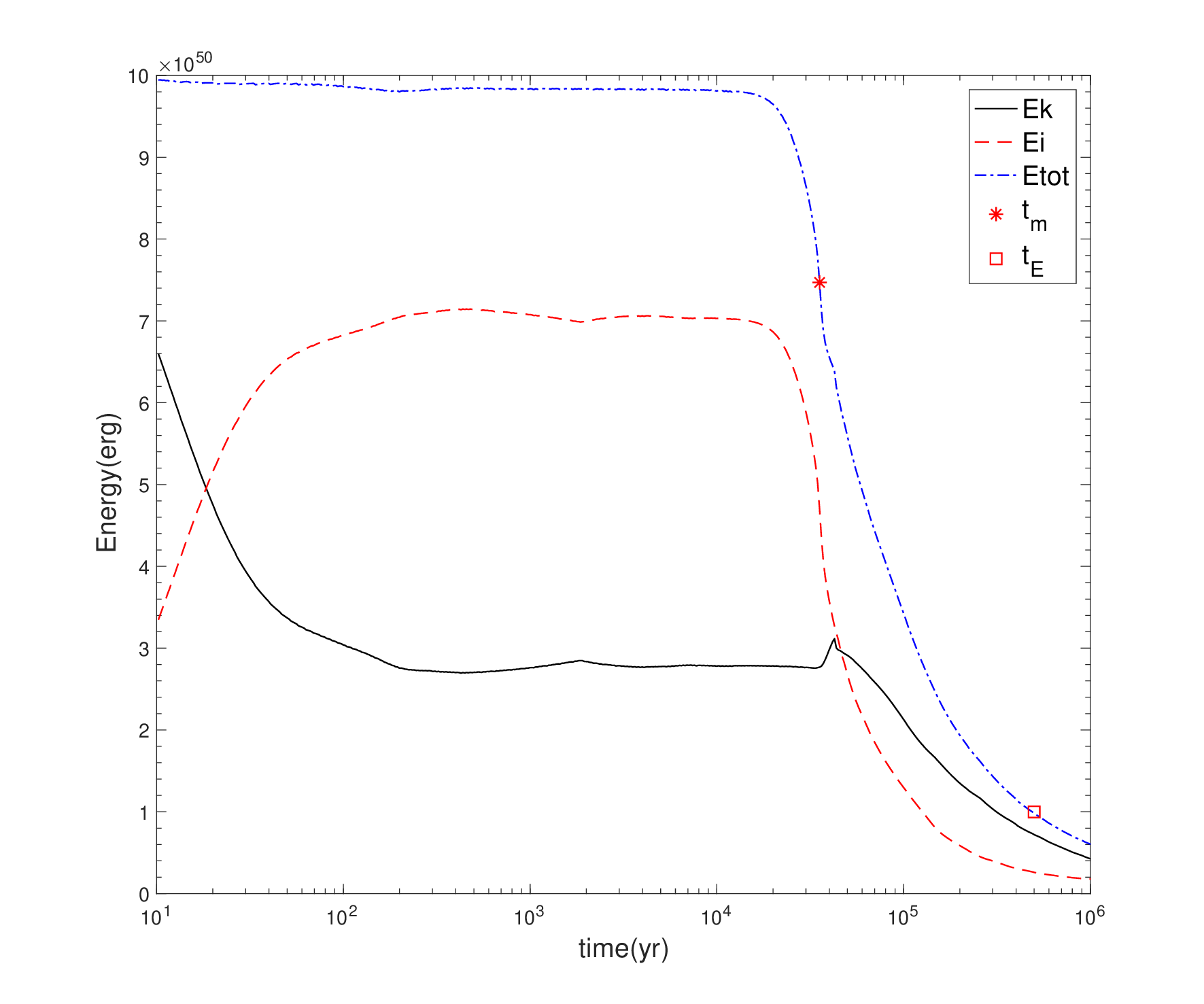}
\caption{The energy of remnant as a function of time.The blue, red,
and black lines are the total energy, internal energy and kinetic
energy, respectively.} \label{fig4}
\end{figure}

\begin{figure}
\centering
\includegraphics[scale=0.8]{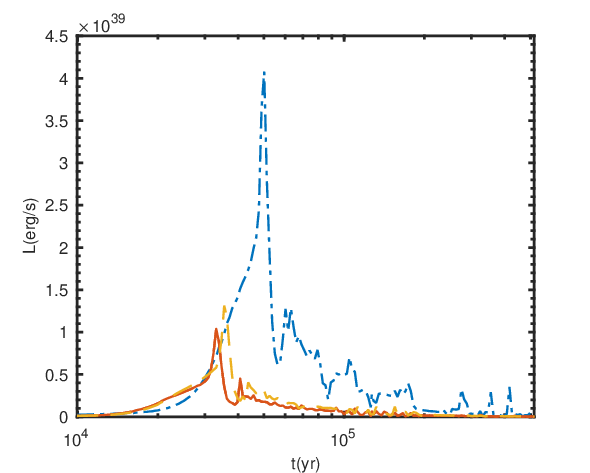}
\caption{The luminosity of remnant as a function of time.The solid,
dashed, and dot-dashed lines are luminosity corresponding to the
ejecta mass of 0.001$M_{\odot}$, 0.05$M_{\odot}$, and
0.1$M_{\odot}$, respectively.} \label{fig5}
\end{figure}

\end{document}